\documentstyle [12pt,epsfig]{article} 
\makeatletter
\@addtoreset{equation}{section}
\makeatother

\newcommand{\ba}{\begin{eqnarray}}
\newcommand{\ea}{\end{eqnarray}}

\begin{document}
\newcommand{\BS}{\bigskip}
\newcommand{\SECTION}[1]{\BS{\large\section{\bf #1}}}
\newcommand{\SUBSECTION}[1]{\BS{\large\subsection{\bf #1}}}
\newcommand{\SUBSUBSECTION}[1]{\BS{\large\subsubsection{\bf #1}}}

\begin{titlepage}
\begin{center}
\vspace*{2cm}
{\large \bf
SPACE-TIME EXCHANGE INVARIANCE: SPECIAL RELATIVITY AS A
SYMMETRY PRINCIPLE}
\vspace*{1.5cm}
\end{center}
\begin{center}
{\bf J.H.Field }
\end{center}
\begin{center}
{ 
D\'{e}partement de Physique Nucl\'{e}aire et Corpusculaire
 Universit\'{e} de Gen\`{e}ve . 24, quai Ernest-Ansermet
 CH-1211 Gen\`{e}ve 4.
}
\end{center}
\vspace*{2cm}
\begin{abstract}
 Special relativity is reformulated as a symmetry property of space-time:
 Space-Time Exchange Invariance.
The additional hypothesis of spatial homogeneity is then sufficient to
derive the Lorentz transformation without reference to the traditional
form of the Principle of Special Relativity. The kinematical version
of the latter is shown to be a consequence of the Lorentz transformation.
As a dynamical application, the laws of
electrodynamics and magnetodynamics are derived from those of 
electrostatics and magnetostatics respectively. The 4-vector nature of the
electromagnetic potential plays a crucial role in the last two
derivations. 
\end{abstract}
\vspace*{4cm}
{\it To be published in American Journal of Physics.}
\end{titlepage}
\section{Introduction}
Two postulates were essential for Einstein's original axiomatic
 derivation~\cite{x1} of the Lorentz transformation (LT) :
 (i) the Special Relativity
Principle and (ii) the hypothesis of the constancy of the velocity of light
in all inertial frames (Einstein's second postulate). The Special Relativity
Principle, which states that:
\par{\bf `The laws of physics are the same in all inertial frames'}
\par had long been known to be respected by Newton's laws of mechanics at the
time Einstein's paper was written. Galileo had already stated the principle 
in 1588 in his `Dialogues Concerning Two New Sciences'. The title of Einstein's
paper~\cite{x1}  `On the Electrodynamics of Moving Bodies' and the special role
of light in his second postulate seem to link special relativity closely to 
classical electrodynamics. Indeed, the LT was discovered as the transformation
that demonstrates that Maxwell's equations may be written in the same way in any
inertial frame, and so manifestly respect the Special Relativity Principle. The same
close connection between special relativity and classical electrodynamics
is retained in virtually all text-book treatments of the subject, obscuring the 
essentially geometrical and kinematical nature of special relativistic effects.
The latter actually transcend the dynamics of any particular physical system.
It was realised, shortly after the space-time geometrical nature of the LT was
pointed out by Minkowski~\cite{x2}, that the domain of applicability of the LT
extends beyond the classical electrodynamics considered by Einstein, and that, in
fact, Einstein's second postulate is not necessary for its derivation~\cite{x3,x4}.
There is now a vast literature devoted to derivations of the LT that do not require
the second postulate~\cite{x5}.
\par In a recent paper by the present author~\cite{x6}, the question of the minimum
number of postulates, in addition to the Special Relativity Principle, necessary to
derive the LT was addressed. The aim of the present paper is somewhat different.
 The Special Relativity Principle itself is re-stated in a simple mathematical form
 which, as will be shown below, has both kinematical and dynamical applications.
 The new statement is a symmetry condition relating space and time, which, it
 is conjectured, is respected by the mathematical
 equations that decscribe all physical laws~\cite{x7}. The symmetry condition
 is first used, together with the postulate of the homogeneity of space, to
 derive the LT. It is then shown that the Kinematical
 Special Relativity Principle (KSRP) is a necessary
 {\it consequence} of the LT. The KSRP, which describes the reciprocal nature of
 similar space time measurements made in two different inertial frames~\cite{x8},
 states that:
 \par{\bf `Reciprocal space-time measurements of similar measuring rods and clocks at
  rest in two different inertial frames $S$, $S'$ by observers at rest in $S'$, $S$ 
  respectively,  yield
  identical results'}
 \par There is no reference here to any physical law. Only space-time events that
 may constitute the raw material of any observation of 
 a physical process are considered. In the previous literature
 the KSRP (or some equivalent condition applied to a gedankenexperiment~\cite{x9})
 has been been used as a necessary postulate to derive the LT.
 \par The symmetry condition that restates the Special Relativity Principle is:
 \begin{itemize} 
 \item[(I)] {\bf `The equations describing the laws of physics are invariant with respect
 to the exchange of space
 and time coordinates, or, more generally, to the exchange of the spatial and temporal 
 components of four vectors.'} 
 \end{itemize}
  A corollary is:
 \begin{itemize}
 \item[(II)] {\bf `Predictions of physical theories do not depend on the metric sign convention
 (space-like or time-like) used to define four-vector scalar products.'}
 \end{itemize}
  A proof of this corollary is presented in Section 4 below.

\par As will become clear during the following discussion, the operation of
 Space-Time Exchange ($STE$) reveals an invariance property of pairs of physical
 equations, which are found to map into each other under $STE$. The examples of this
 discussed below are: the Lorentz transformation equations of space and time, the
 Maxwell equations describing electrostatics (Gauss' law) and electrodynamics
 (Amp\`{e}re's law), and those describing magnetostatics (Gauss' law) and magnetodynamics
 (The Faraday-Lenz law). It will be demonstrated
 that each of these three pairs of equations map into each other under $STE$, and so
 are invariants of the $STE$ operator. In the case of the LT equations, imposing $STE$ symmetry
 is sufficient to derive them from a general form of the space transformation
 equation that respects the classical limit.

\par The expression: `The equations describing the laws of physics' in (I) should then be
 understood as including {\it both equations} of each $STE$ invariant pair. For example, the 
 Gauss equation of electrostatics, considered as an independent physical law, clearly does
 not respect (I).

\par For dimensional reasons, the definition of the exchange operation
referred to in (I) requires the time coordinate to be multiplied by a universal 
parameter $V$ with the dimensions of velocity. The new time coordinate with
dimension$[L]$: 
\begin{equation}
 x^0  \equiv Vt
 \end{equation}
 may be called the `causality radius'~\cite{x10} to distinguish it from the 
 cartesian spatial coordinate $x$ or the invariant interval $s$. Since space
 is three dimensional and time is one dimensional, there is a certain ambiguity in
 the definition of the exchange operation in (I). Depending on the case under
 discussion, the space coordinate may be either the magnitude of the spatial vector
 $x=|\vec{x}|$, or a cartesian component $x^1$,$x^2$,$x^3$. For any physical
 problem with a preferred spatial direction (which is the case for the LT), then,
 by a suitable choice of coordinate system, the identification $x=x^1$, $x^2=x^3=0$
 is always possible.  The exchange
 operation in (I) is then simply $x^0 \leftrightarrow x^1$. Formally, the exchange
 operation is defined by the equations:
 \begin{eqnarray}
     STE x^0 = x^1  \\
     STE x^1 = x^0   \\
     (STE)^2 = 1
 \end{eqnarray}
 where  $STE$ denotes the space time exchange operator. As shown below, for problems
 where there is no preferred direction, but rather spatial symmetry, it may also
 be useful to define three exchange operators:
\begin{equation}
  x^0 \leftrightarrow x^i~~~~i=1,2,3
\end{equation}
 with associated operations $STE(i)$ analagous to $STE=STE(1)$ in Eqns.(1.2)-(1.4).
 The operations in Eqns.(1.2) to (1.5) may also be generalised to the case of an 
 arbitary 4-vector with temporal and spatial components $A^0$ and $A^1$ respectively.
\par To clarify the meaning of the $STE$ operation, it is of interest to compare it with
 a different operator acting on space and time coordinates that may be called
 `Space-Time Coordinate Permutation' ($STCP$).
 Consider an equation of the form:
\begin{equation}
  f(x^0,x^1)=0.    
\end{equation}
 The $STE$ conjugate equation is:
\begin{equation}
  f(x^1,x^0)=0.    
\end{equation} 
This equation is different from (1.6) because $x^0$ and $x^1$ have different
physical meanings. In the $STCP$ operation however, the {\it values} of the 
space and time coordinates are interchanged, but no new equation is generated.
If $x^0=a$ and $x^1=b$ in Eqn.(1.6) then the $STCP$ operation applied to the 
latter yields:
\begin{equation}
  f(x^0=b ,x^1=a)=0.    
\end{equation}
 This equation is identical in form to (1.6); only its parameters have
 different values.

 \par The physical meaning of the universal parameter $V$, and its relation to the
 velocity of light, $c$, is discussed in the following Section, after the derivation
 of the LT. 
 \par The plan of the paper is as follows. In the following Section the LT is derived.
 In Section 3, the LT is used to derive the KSRP. The space time exchange properties
 of 4-vectors and the related symmetries in Minkowski space are discussed in Section 4.
 In Section 5 the space-time exchange symmetries of Maxwell's equations are used to
 derive electrodynamics (Amp\`{e}re's law) and magnetodynamics (the Faraday-Lenz
 law) from the Gauss laws of electrostatics and magnetostatics respectively. A summary is
 given in Section 6.
  
\section{Derivation of the Lorentz Transformation}
 Consider two inertial frames $S$,$S'$. $S'$ moves along the common $x,x'$ axis of orthogonal 
 cartesian coordinate systems in $S$,$S'$ with velocity $v$ relative to $S$. The $y,y'$ axes 
 are also parallel. At time $t=t'=0$ the origins of $S$ and $S'$ coincide. In general the
 transformation equation between the coordinate $x$ in $S$ of a fixed point on the 
 $Ox'$ axis and the coordinate $x'$ of the same point referred to the frame $S'$ is :
 \begin{equation}
 x' = f(x,x^0,\beta)
 \end{equation}
 where $ \beta \equiv v/V$ and $V$ is the universal constant introduced in Eqn.(1.1).
 Differentiating Eqn.(2.1) with respect to $x^0$, for fixed $x'$, gives:
 \begin{equation}
 \left. \frac{dx'}{dx^0} \right|_{x'} = 0 = \left.\frac{dx}{dx^0} \right|_{x'}
 \frac{\partial f}{\partial x}+\frac{ \partial f}{\partial x^0}
 \end{equation}
 Since 
 \[ \left.\frac{dx}{dx^0} \right|_{x'} = \frac{1}{V} \left.\frac{dx}{dt} \right|_{x'}
 = \frac{v}{V} = \beta\]
 the function $f$ must satisfy the  partial differential equation:
 \begin{equation}
 \beta \frac{\partial f}{\partial x} = -\frac{ \partial f}{\partial x^0}
 \end{equation}
 A sufficient condition for $f$ to be a solution of Eqn.(2.3) is that it is a function
 of $x-\beta x^0$. Assuming also $f$ is a differentiable function, it may be expanded
 in a Taylor series:
 \begin{equation}
 x' = \gamma (\beta)(x-\beta x^0) + \sum_{n=2}^{\infty} a_n(\beta)(x-\beta x^0)^n
 \end{equation}
 Requiring either spatial homogeneity~\cite{x11,x12,x13}, or that the LT is a unique, single valued,
  function of its arguments~\cite{x6}, requires Eqn.(2.4) to be linear, i.e.
  \[ a_2(\beta) = a_3(\beta) = .~.~. = 0 \]
  so that
 \begin{equation}
 x' = \gamma (\beta)(x-\beta x^0) 
 \end{equation}
 Spatial homogeneity implies that Eqn(2.5) is invariant when all spatial coordinates
 are scaled by any constant factor $K$. Noting that :
 \begin{equation}
 -\beta = - \frac{1}{V} \left.\frac{dx}{dt} \right|_{x'} 
  = \frac{1}{V} \left.\frac{d(-x)}{dt} \right|_{x'}  
 \end{equation}     
 and choosing $K=-1$ gives :
 \begin{equation}
 -x' = \gamma (-\beta)(-x+\beta x^0) 
 \end{equation} 
 Hence, Eqn.(2.5) is invariant provided that 
 \begin{equation}
 \gamma(-\beta) = \gamma(\beta)
 \end{equation}
 i.e. $\gamma(\beta)$ is an even function of $\beta$. 
\par Applying the space time exchange operations $x \leftrightarrow x^0$, 
 $ x' \leftrightarrow (x^0)'$ to Eqn.(2.5) gives
\begin{equation}
(x^0)' = \gamma (\beta)(x^0-\beta x) 
 \end{equation}  
 The transformation inverse to (2.9) may, in general, be written as:
\begin{equation}
 x^0 = \gamma (\beta')((x^0)'-\beta' x') 
 \end{equation}       
 The same inverse transformation may also be derived by eliminating $x$
 between Eqns.(2.5) and (2.9) and re-arranging:
\begin{equation}
 x^0 = \frac{1}{\gamma(\beta)(1-\beta^2)}((x^0)'+\beta x') 
 \end{equation}   
 Eqns(2.10),(2.11) are consistent provided that :
\begin{equation}
\gamma (\beta')  = \frac{1}{\gamma(\beta)(1-\beta^2)} 
 \end{equation}    
 and 
\begin{equation}
 \beta' =  -\beta 
 \end{equation}  
 Eqns.(2.8),(2.12) and (2.13) then give~\cite{x14}: 
\begin{equation}
 \gamma(\beta) =  \frac{1}{\sqrt{1-\beta^2}} 
 \end{equation}      
 Eqns.(2.5),(2.9) with $\gamma$ given by (2.14) are the LT equations for space-time
 points along the common $x,x'$ axis of the frames $S$,$S'$. They have been derived here
 solely from the symmetry condition (I) and the assumption of spatial homogeneity,
 without any reference to the Principle of Special Relativity. 
 \par The physical meaning of the universal parameter $V$ becomes clear when the 
 kinematical consequences of the LT for physical objects are worked out in detail.
 This is done, for example, in Reference [6], where it is shown that the velocity of any
 massive physical object approaches $V$ in any inertial frame in which its energy is
 much greater than its rest mass. The identification of $V$ with the velocity of light,
 $c$, then follows~\cite{x13,x6} if it is assumed that light consists of massless
 (or almost massless) particles, the light quanta discovered by Einstein in his
 analysis of the photoelectric effect~\cite{x15}. That $V$ is the limiting velocity
 for the applicability of the LT equations is, however, already evident from Eqn.(2.14).
 If $\gamma(\beta)$ is real then $\beta \le 1$, that is $v \le V$.

\section{Derivation of the Kinematical Special Relativity Principle}
The LT equations (2.5) and (2.9) and their inverses, written in terms of $x,x';t,t'$ are:
\begin{eqnarray}
x'= \gamma (x-vt) \\
t'= \gamma (t-\frac{vx}{V^2}) \\
x= \gamma (x'+vt') \\
t= \gamma (t'+\frac{vx'}{V^2}) 
\end{eqnarray}
Consider now observers, at rest in the frames $S$,$S'$, equipped with identical measuring
rods and clocks. The observer in $S'$ places a rod, of length $l$, along the common $x,x'$
axis. The coordinates in $S'$ of the ends of the rod are $x_1',x_2'$ where 
$x_2'-x_1'=l$. If the observer in $S$ measures, at time $t$ in his own frame, the ends
of the rod to be at $x_1, x_2$ then, according to Eqn(3.1):
\begin{eqnarray}
x_1'= \gamma (x_1-vt) \\
x_2'= \gamma (x_2-vt)
\end{eqnarray} 
Denoting by $l_S$ the apparent length of the rod, as observed from $S$ at time $t$,
 Eqns.(3.5),(3.6) give
\begin{equation}
l_S \equiv x_2-x_1 = \frac{1}{\gamma}(x_1'-x_2') = \frac{l}{\gamma}
\end{equation} 
Suppose that the observer in $S'$ now makes reciprocal measurements $x_1',x_2'$ of the
ends of a similar rod, at rest in $S$, at time $t'$. In $S$ the ends of the rod are at
the points $x_1,x_2$, where $l=x_2-x_1$. Using Eqn.(3.3)
\begin{eqnarray}
x_1= \gamma (x_1'+vt') \\
x_2= \gamma (x_2'+vt')
\end{eqnarray}   
and, corresponding to (3.7), there is the relation:
\begin{equation}
l_{S'} \equiv x_2'-x_1' = \frac{1}{\gamma}(x_2-x_1) = \frac{l}{\gamma}
\end{equation} 
Hence, from Eqns.(3.7),(3.10)
\begin{equation}
l_{S} = l_{S'} = \frac{l}{\gamma}
\end{equation}
so that reciprocal length measurements yield identical results.
\par Consider now a clock at rest in $S'$ at $x'=0$.
This clock is synchronized with a similar clock in $S$ at $t=t'=0$, when the
spatial coordinate systems in $S$ and $S'$ coincide. Suppose that the observer at rest in
$S$ notes the time $t$ recorded by his own clock, when the moving clock records the time
$\tau$. At this time, the clock which is moving along the common $x,x'$ axis with
 velocity $v$ will be situated at $x=vt$. With the definition $\tau_S \equiv t$, and using
 Eqn.(3.2) : 
\begin{equation}
\tau = \gamma (\tau_S-\frac{vx}{V^2})=\gamma \tau_S (1-\frac{v^2}{V^2}) =
\frac{\tau_S}{\gamma}
\end{equation}    
If the observer at rest in $S'$ makes a reciprocal measurement of the clock at rest in
$S$, which is seen to be at $x' =-vt'$ when it shows the time $\tau$, then according
to Eqn.(3.4) with $\tau_{S'} \equiv t'$: 
\begin{equation}
\tau = \gamma (\tau_{S'}+\frac{vx'}{V^2})=\gamma \tau_{S'} (1-\frac{v^2}{V^2}) =
\frac{\tau_{S'}}{\gamma} 
\end{equation}    
Eqns.(3.12),(3.13) give
\begin{equation}
\tau_{S} = \tau_{S'} = \gamma \tau
\end{equation}
Eqns.(3.11),(3.14) prove the Kinematical Special Relativity Principle as stated above.
It is a necessary consequence of the LT.

\section{General Space Time Exchange Symmetry Properties of 4-Vectors.
Symmetries of Minkowski Space} 
The LT was derived above for space time points lying along the common $x,x'$
axis, so that $x = |\vec{x}|$. However, this restriction is not necessary.
In the case that $\vec{x}=(x^1,x^2,x^3)$ then $x$ and $x'$ in Eqn.(2.5) may
be replaced by $x = \vec{x} \cdot \vec{v}/|\vec{v}|$ and 
$x' = \vec{x'} \cdot \vec{v}/|\vec{v}|$ respectively, where the 1-axis
is chosen parallel to $\vec{v}$. The proof proceeds as before with
the space time exchange operation defined as in Eqns.(1.2)-(1.4). The
additional transformation equations :
\begin{eqnarray}
 y'=y  \\
 z'=z 
\end{eqnarray}
follow from spatial isotropy~\cite{x1}.
\par In the above derivation of the LT, application of the $STE$ operator generates
the LT of time from that of space. It is the pair of equations that is 
invariant with respect to the $STE$ operation. Alternatively, as shown below,
by a suitable change of variables, equivalent equations may be defined that are
manifestly invariant under the $STE$ operation.  
\par The 4-vector velocity $U$  and the energy-momentum 4-vector $P$ are defined in terms of the
space-time 4-vector~\cite{x2}:
\begin{equation}
X \equiv (Vt;x,y,z) = (x^0;x^1,x^2,x^3)
\end{equation}
by the equations:
\begin{eqnarray}
 U \equiv \frac{dX}{d \tau}  \\
  P \equiv mU 
\end{eqnarray}
where $m$ is the Newtonian mass of the physical object and $\tau$ is its proper 
time, i.e. the time in a reference frame in which the object is at rest.
Since $\tau$ is a Lorentz invariant quantity, the 4-vectors $U,P$ have identical 
LT properties to X. The properties of $U,P$ under the $STE$ operation follow directly
from Eqns.(1.2),(1.3) and the definitions (4.4) and (4.5).
Writing the energy-momentum 4-vector as:
\begin{equation}
P = (\frac{E}{V}; p,0,0) = (p^0;p^1,0,0)
\end{equation}  
the $STE$ operations: $p^0 \leftrightarrow p^1$ , $(p^0)' \leftrightarrow (p^1)'$
generate the LT equation for energy:
\begin{equation}
(p^0)'= \gamma ( p^0-\beta p^1)
\end{equation}
from that of momentum
\begin{equation}
(p^1)'= \gamma ( p^1-\beta p^0)
\end{equation}
or {\it vice versa}.
\par The scalar product of two arbitary 4-vectors $C$,$D$:
\begin{equation}
C \cdot D \equiv C^0 D^0-\vec{C} \cdot \vec{D}
\end{equation}
can, by choosing the x-axis parallel to $\vec{C}$ or $\vec{D}$, always be written as:
\begin{equation}
C \cdot D = C^0 D^0-C^1 D^1
\end{equation}  
Defining the $STE$ exchange operation for an arbitary 4-vector in a similar
way to Eqns.(1.2),(1.3) then the combined operations $C^0 \leftrightarrow C^1$,
$D^0 \leftrightarrow D^1$ yield:
\begin{equation}
 C \cdot D \rightarrow  C^1 D^1-C^0 D^0 = - C \cdot D
\end{equation}  
 The 4-vector product changes sign, and so the combined $STE$ operation is equivalent
 to a change in the sign convention of the metric from space-like to time-like (or
 {\it vice versa} ), hence the corollary (II) in Section 1 above.

\par The LT equations take a particularly simple form if new variables are
 defined which have simple transformation properties under the $STE$ operation.
 The variables are:
 \begin{eqnarray}
 x_+ = \frac{x^0+x^1}{\sqrt{2}} \\
 x_- = \frac{x^0-x^1}{\sqrt{2}}
 \end{eqnarray}
$x_+$, $x_-$ have, respectively, even and odd `$STE$ parity':
 \begin{eqnarray}
STE x_+ = x_+  \\
STE x_- = -x_-
 \end{eqnarray}
 The manifestly $STE$ invariant LT equations expressed in terms of these
 variables are:
 \begin{eqnarray}
 x'_+ = \alpha x_+  \\
 x'_- = \frac{1}{\alpha}x_-
 \end{eqnarray}  
 where
 \begin{equation}
 \alpha = \sqrt{\frac{1-\beta}{1+\beta}}
 \end{equation}
 Introducing similar variables for an arbitary 4-vector:
 \begin{eqnarray}
 C_+ = \frac{C^0+C^1}{\sqrt{2}}  \\
 C_- = \frac{C^0-C^1}{\sqrt{2}}
 \end{eqnarray} 
 the 4-vector scalar product of $C$ and $D$ may be written as:
 \begin{equation}
 C \cdot D = C_+D_- + C_-D_+
 \end{equation}

\begin{figure}[htbp]
\begin{center}\hspace*{-0.5cm}\mbox{
\epsfysize15.0cm\epsffile{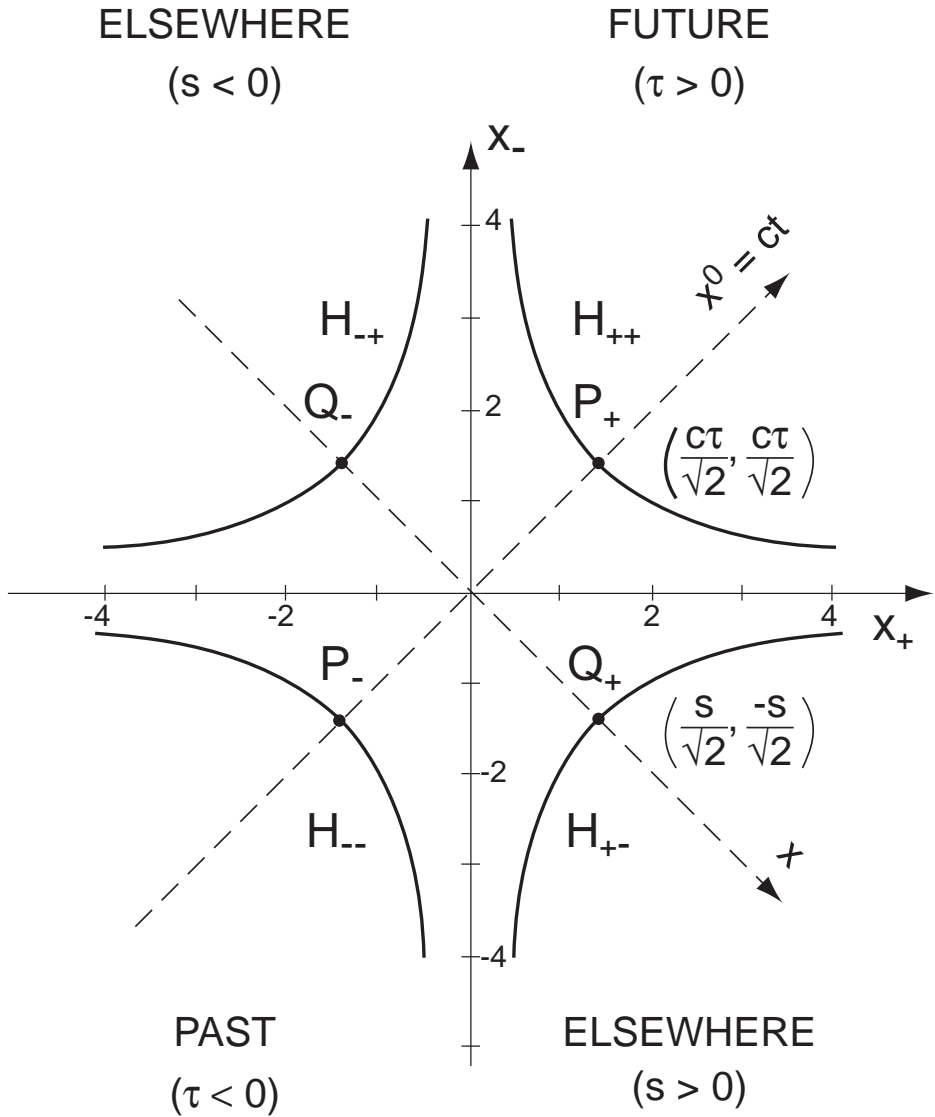}}
\caption{Space-time points in S' as seen by an observer in S. The
hyperbolae $H_{++}$, $H_{--}$ correspond to points at the origin of S' at
time $t' = \tau$. The
hyperbolae $H_{+-}$, $H_{-+}$ correspond to points at $x' = s$ and $t' = 0$.
See the text for the equations of the hyperbolae and further discussion.}
\end{center}
\end{figure}

 In view of the LT equations (4.16),(4.17) $C \cdot D$ is manifestly Lorentz
 invariant. The transformations (4.12),(4.13) and (4.19),(4.20) correspond
 to an anti-clockwise rotation by $45^{\circ}$ of the axes of the usual
  $ct$ versus $x$ plot. The $x_+$,$x_-$ axes lie along the light cones
 of the $x$-$ct$ plot (see Fig.1). 
 \par The LT equations (4.16),(4.17) give a parametric
 representation of a hyperbola in $x_+$,$x_-$ space. A point on the latter
 corresponds to a particular space-time point as viewed in a frame $S$. 
  The point $x_+ =x_-=0$ corresponds to the space-time origin of the frame
  $S'$ moving with velocity  $\beta c$ relative to $S$. A point at the spatial
   origin
  of $S'$ at time $t' =\tau$ will be seen by an observer in $S$,
  as $\beta$ (and hence $\alpha$) varies, to lie on one of the hyperbolae
 $H_{++}$, $H_{--}$ in Fig.1:
  \begin{equation}
  x_+x_-= \frac{c^2 \tau^2}{2}
  \end{equation}
  with $x_+,x_- > 0$ if $\tau > 0$ ($H_{++}$) or $x_+,x_- < 0$ if $\tau < 0$
 ($H_{--}$). A point 
  along the $x'$ axis at a distance $s$ from the origin, at $t' =0$ lies on
  the hyperbolae $H_{+-}$, $H_{-+}$:
  \begin{equation}
  x_+x_-= -\frac{s^2}{2}
  \end{equation}
 with $x_+>0$, $x_-< 0$ if $s > 0$ ($H_{+-}$)  or $x_+<0$, $x_->0$ if
 $s < 0$ ($H_{-+}$).    
  As indicated in Fig.1 the hyperbolae (4.22)
  correspond to the past ($\tau < 0$) or the
  future ($\tau > 0$) of a space time point at the origin of $S$ or $S'$, whereas
  (4.23) corresponds to the `elsewhere' of the same space-time points. That is, the
   manifold of all space-time points that are causally disconnected from them.
   These are all familiar properties of the Minkowski space x-ct plot. One 
   may note, however, the simplicity of the equations (4.16),(4.17),(4.22),
   (4.23) containing the `lightcone' variables $x_+,x_-$ that have simple
   transformation properties under the $STE$ operation.
  \par Another application of $STE$ symmetry may be found in~\cite{x17}.
  It is shown there that the apparent distortions of space-time that occur
  in observations of moving bodies or clocks are related by this symmetry.
  For example, the Lorentz-Fitzgerald contraction is directly related to
  Time Dilatation by the $STE$ operations (1.2) and (1.3).
  
\section{Dynamical Applications of Space Time Exchange Symmetry}
If a physical quantity is written in a manifestly covariant way, as a function
of 4-vector products, it will evidently be invariant with respect to $STE$ as 
the exchange operation has the effect only of changing the sign convention for 
4-vector products from space-like to time-like or {\it vice-versa}. An example
of such a quantity is the invariant amplitude $\cal M$ for an arbitary 
scattering process in Quantum Field Theory. In this case $STE$ invariance is 
equivalent to Corollary II of Section 1 above.
\par More interesting results can be obtained from equations where components
 of 4-vectors appear directly. It will now be shown how $STE$ invariance may be
 used to derive Amp\`{e}re's law and Maxwell's `displacement current' from the
 Gauss law of electrostatics, and the Faraday-Lenz law of magnetic induction
 from the the Gauss law of magnetostatics (the absence of magnetic charges).
 Thus electrodynamics and magnetodynamics follow from the laws of 
 electrostatics and magnetostatics, together with space time exchange symmetry 
 invariance. It will be seen that the 4-vector character of the electromagnetic
 potential plays a crucial role in these derivations.
 \par In the following, Maxwell's equations are written in Heaviside-Lorentz
 units with $V = c = 1$~\cite{x18}. The 4-vector potential $A = (A^0;\vec{A})$
 is related to the electromagnetic field tensor $F^{\mu \nu}$ by the equation:
 \begin{equation}
  F^{\mu \nu} = \partial^{\mu} A^{\nu} - \partial^{\nu} A^{\mu}
 \end{equation}
 where
 \begin{equation}
 \partial^{\mu} \equiv  (\frac{\partial~}{\partial t} ; -\vec{\nabla} ) =
 (\partial^0 ; -\vec{\nabla})
 \end{equation}
 The electric and magnetic field components $E^k$, $B^k$ respectively, are 
 given, in terms of $F^{\mu \nu}$, by the equations:
 \begin{eqnarray}
   E^k & = & F^{k0} \\
   B^k & = & -\epsilon_{ijk} F^{ij}
 \end{eqnarray}
 A time-like metric is used with $C_t = C^0 = C_0$, $C_x = C^1 = - C_1$ etc,
 with summation over repeated contravariant (upper) and covariant (lower)
 indices understood. Repeated greek indices are summed form 1 to 4 and roman ones from
 1 to 3.
\par The transformation properties of contravariant and covariant 4-vectors under the
 $STE$ operation are now discussed. They are derived from the general condition that 
4-vector products change sign under the $STE$ operation (Eqn.(4.11)). The 4-vector
 product (4.10) is written, in terms of contravariant and covariant 4-vectors, as:
\begin{equation}
C \cdot D = C^0D_0+C^1D_1
\end{equation}
Assuming that the contravariant 4-vector $C^{\mu}$ transforms according to Eqns.(1.2)
(1.3), i.e.
\begin{equation}
 C^0 \leftrightarrow C^1
\end{equation}
the covariant 4-vector $D_{\mu}$ must transform as:
\begin{equation}
 D_0 \leftrightarrow -D_1
\end{equation} 
in order to respect the transformation property 
\begin{equation}
 C \cdot D \rightarrow -C \cdot D 
\end{equation}
of 4-vector products under $STE$.

\par It remains to discuss the  $STE$ transformation properties of $\partial^{\mu}$
and the 4-vector potential $A^{\mu}$. In view of the property of $\partial^{\mu}$:
$\partial^1= -\partial_x=-\partial/\partial x$ (Eqn.(5.2)), which is similar to the
relation $C_1 = -C_x$ for a {\it covariant} 4-vector, it is natural to choose for 
 $\partial^{\mu}$ an $STE$ transformation similar to Eqn.(5.7):
\begin{equation}
\partial^0 \leftrightarrow  -\partial^1
\end{equation}
and hence, in order that $\partial^{\mu}\partial_{\mu}$ change sign under $STE$:
\begin{equation}
\partial_0 \leftrightarrow  \partial_1
\end{equation}
\par This is because it is clear that the appearence of a minus sign in the 
$STE$ transformation equation (5.7) is correlated to the minus sign in
 front of the spatial components of a covariant 4-vector, not whether
 the Lorentz index is an upper or lower one. Thus $\partial^{\mu}$ and
 $\partial_{\mu}$  transform in an `anomalous' manner under $STE$ as
 compared to the convention of Eqns.(5.6) and (5.7). In order that
 the 4-vector product  $\partial_{\mu}A^{\mu}$ respect the condition (5.8),
$A^{\mu}$ and $A_{\mu}$ must then transform under $STE$ as:
\begin{equation}
A^0 \leftrightarrow  -A^1
\end{equation}
and 
\begin{equation}
A_0 \leftrightarrow  A_1
\end{equation}
respectively. That is, they transform in the same way as
 $\partial^{\mu}$ and $\partial_{\mu}$ respectively.

\par Introducing the 4-vector electromagnetic current
   $j^{\mu} \equiv (\rho;\vec{j})$, Gauss' law of electrostatics may be
    written as:
  \begin{equation}
  \vec{\nabla} \cdot \vec{E} = \rho = j^0
  \end{equation}
  or, in the manifestly covariant form:
  \begin{equation}
  (\partial_{\mu} \partial^{\mu})A^0-\partial^0(\partial_{\mu} A^{\mu})= j^0
   \end{equation}
   This equation is obtained by writing Eqn.(5.13) in covariant notation 
   using Eqns.(5.1) and (5.3) and adding to the left side the identity:
  \begin{equation}
  \partial_0 (\partial^0 A^0- \partial^0 A^0) = 0
   \end{equation}      
Applying the space-time exchange operation to Eqn.(5.14), with 
index exchange $ 0 \rightarrow 1 $ (noting that 
$\partial^0$, $A^0$ transform according to Eqns(5.9),(5.11), $j^0$
according to (5.6), and that the
scalar products $\partial_{\mu} \partial^{\mu}$ and  $\partial_{\mu} A^{\mu}$
change sign) yields the equation:
  \begin{equation}
  (\partial_{\mu} \partial^{\mu})A^1-\partial^1(\partial_{\mu} A^{\mu})= j^1
   \end{equation}        
  The spatial part of the 4-vector products on the left side of Eqn.(5.16) is:
\begin{eqnarray}
 \partial_i (\partial^i A^1- \partial^1 A^i) & = & \partial_i F^{i1}
 \nonumber \\ 
         & = & \partial_2 B^3- \partial_3 B^2 \nonumber \\
         & = &  (\vec{\nabla} \times \vec{B} )^1
\end{eqnarray}
where Eqns.(5.1) and (5.4) have been used.
 The time part of the 4-vector products in Eqn(5.16) yields, with Eqns.(5.1) 
 and (5.3):
 \begin{equation}
\partial_0 (\partial^0 A^1- \partial^1 A^0) =  -\frac{\partial E^1}{\partial t}
 \end{equation}
 Combining Eqns(5.16)-(5.18) gives:
 \begin{equation} 
 (\vec{\nabla} \times \vec{B} )^1 -\frac{\partial E^1}{\partial t} = j^1
 \end{equation}
 Combining Eqn.(5.19) with the two similar equations derived derived by the 
 index exchanges $0 \rightarrow 2$, $ 0 \rightarrow 3$ in Eqn.(5.14) gives:
 \begin{equation} 
 (\vec{\nabla} \times \vec{B} ) -\frac{\partial\vec{E}}{\partial t} = \vec{j}
 \end{equation} 
 This is Amp\`{e}re's law, together with Maxwell's displacement current.
 \par The Faraday-Lenz law is now derived by applying the space-time exchange
 operation to the Gauss law of magnetostatics:
 \begin{equation}
 \vec{\nabla} \cdot \vec{B} = 0
 \end{equation}      
 Introducing Eqns.(5.4) and (5.1) into Eqn.(5.21) gives:
 \begin{equation}
\partial_1 (\partial^3 A^2- \partial^2 A^3)+
\partial_2 (\partial^1 A^3- \partial^3 A^1)+
\partial_3 (\partial^2 A^1- \partial^1 A^2) = 0
 \end{equation}
Making the exchange $1 \rightarrow 0$ of space-time indices in Eqn.(5.22) and
noting that $\partial_1$ transforms according to Eqn.(5.10), whereas $\partial^1$,
$A^1$  transform as in Eqns.(5.9),(5.11) respectively, gives:
 \begin{equation}
\partial_0 (\partial^3 A^2- \partial^2 A^3)+
\partial_2 (-\partial^0 A^3+ \partial^3 A^0)+
\partial_3 (-\partial^2 A^0- \partial^0 A^2) = 0
 \end{equation}
 Using Eqns.(5.1)-(5.4), Eqn.(5.23) may be written as:
 \begin{equation}
 \frac{\partial B^1}{\partial t} +\partial_2 E^3-\partial_3 E^2 = 0
  \end{equation} 
  or, in 3-vector notation:
\begin{equation} 
 (\vec{\nabla} \times \vec{E} )^1 = -\frac{\partial B^1}{\partial t} 
 \end{equation}                
 The space-time exchanges $2 \rightarrow 0$, $3 \rightarrow 0$ in Eqn.(5.22)
 yield, in a similar manner, the 2 and 3 components of the Faraday-Lenz law:
 \begin{equation} 
 (\vec{\nabla} \times \vec{E} ) = -\frac{\partial\vec{B}}{\partial t}
 \end{equation}  
 \par Some comments now on the conditions for the validity of the above
 derivations. It is essential to use the manifestly covariant
 form of the electrostatic Gauss law Eqn.(5.14) and the manifestly rotationally
 invariant form, Eqn.(5.22), of the magnetostatic Gauss law. For example,
 the  1-axis may be chosen
 parallel to the electric field in Eqn.(5.13). In this case Eqn.(5.14) simplifies
 to 
   \begin{equation}
  \partial_1 (\partial^0 A^1- \partial^1 A^0) = j^0
   \end{equation}
   Applying the space-time exchange operation $0 \leftrightarrow 1$ to this
   equation yields only the Maxwell displacement current term in Eqn.(5.19).
   Similarly, choosing the 1-axis parallel to $\vec{B}$ in Eqn.(5.21)
   simplifies Eqn.(5.22) to 
   \begin{equation}
  \partial_1 (\partial^3 A^2- \partial^2 A^3) = 0
   \end{equation}   
  The index exchange $1 \rightarrow 0$ leads then to the equation:
 \begin{equation}  
 \frac{\partial B^1}{\partial t} = 0 
  \end{equation}
  instead of the 1-component of the Faraday-Lenz law, as in Eqn.(5.24).
\par The choice of the $STE$ transformation properties of contravariant 
 and covariant 4-vectors according to Eqns.(5.6) and (5.7) is an arbitary
 one. Identical results are obtained if the opposite convention is used.
 However, `anomalous' transformation properties of $\partial^{\mu}$,
 $\partial_{\mu}$ and $A^{\mu}$, $A_{\mu}$, in the sense described 
 above, are essential. This complication results from the upper
index on the left side of Eqn.(5.2) whereas on the right side
 the spatial derivative is multiplied by a minus sign. This minus
 sign changes the $STE$ transformation property relative to that, (5.6),
 of conventional contravariant 4-vectors, that do not have a minus
 sign multiplying the spatial components. The upper index on the 
 left side of Eqn.(5.2) is a consequence of the Lorentz transformation
 properties 
 of the four dimensional space-time derivative~\cite{Weinberg}.

\section{Summary and Discussion} 
In this paper the Lorentz transformation for points lying along the common
$x$, $x'$ axis of two inertial frames has been derived from only
 two postulates: (i) the symmetry principle (I), and (ii) the homogeneity
  of space.
 This is the same number of axioms as used in Ref.[6] where
  the postulates were: the
 Kinematical Special Relativity Postulate and the uniqueness condition.
 Since both spatial homogeneity and uniqueness require the LT equations to be 
 linear, the KSRP of Ref.[6] has here, essentially, been replaced by the
 space-time symmetry condition (I).
 \par Although postulate (I) and the KRSP play equivalent roles in the
 derivation of the LT, they state in a very different way the physical
 foundation of special relativity. Postulate (I) is a mathematical statement
 about the structure of the equations of physics, whereas the KSRP makes,
 instead, a statement about the relation between space-time measurements
 performed in two different inertial frames. It is important to note that
 in neither case do the dynamical laws describing any particular physical
 phenomenon enter into the derivation of the LT.
 
 \par Choosing postulate (I) as the fundamental principle of special relativity
 instead of the Galilean Relativity Principle, as in the traditional approach,
 has the advantage that a clear distinction is made, from the outset, between
 classical and relativistic mechanics. Both the former and the latter respect
 the Galilean Relativity Principle but with different laws. On the other hand, 
 only relativistic equations, such as the LT or Maxwell's Equations, respect
 the symmetry condition (I).
 \par The teaching of, and hence the understanding of, special relativity 
 differs greatly depending on how the parameter $V$ is introduced.
 In axiomatic derivations of the LT, that do not use Einstein's second
  postulate, a universal parameter $V$ with the dimensions of velocity 
  necessarily appears at an intermediate stage of the derivation~\cite{x19}.
  Its physical meaning, as the absolute upper limit of the observed velocity
  of {\it any} physical object, only becomes clear on working out the
   kinematical consequences of the LT~\cite{x6}. If Einstein's second
   postulate is
   used to introduce the parameter $c$, as is done in the vast majority
   of text-book treatments of special relativity, justified by the
   empirical observation of the constancy of the velocity of light, the
   actual universality of the theory is not evident. The misleading impression
   may be given that special relativity is an aspect of classical electrodynamics,
   the domain of physics in which it was discovered.
   \par Formulating special relativity according to the symmetry principle (I)
   makes clear
   the space-time geometrical basis~\cite{x2} of the theory. The universal 
   velocity parameter $V$ must be introduced at the outset in order even
   to define
   the space-time exchange operation. Unlike the Galilean Relativity Principle,
 the symmetry condition (I) gives a clear test of 
  whether any physical equation is a candidate to describe a universal law of 
  physics. Such an equation must either be invariant under space-time exchange
  or related by the exchange operation to another equation that also represents
  a universal law. The invariant amplitudes of quantum field theory are an example
 of the former case, while the LT equations for space and time correspond to the
 latter. Maxwell's equations are examples of dynamical laws that satisfy the
  symmetry condition (I). The laws of electrostatics and magnetostatics (Gauss'
  law for electric and magnetic charges) are related by the space-time
  exchange symmetry to the laws of electrodynamics (Amp\`{e}re's law)
     and magnetodynamics (the Faraday-Lenz law) respectively.
   The 4-vector character~\cite{x20} of the
   electromagnetic potential is essential for these symmetry
    relations~\cite{x21}.
\par \underline{Acknowledgement}

\par  I thank an anonymous referee for his encouragement, as well
 as for many suggestions that have enabled me to much improve the clarity of
 the presentation. The assistance of C.Laignel in the preparation of the 
 figure is also gratefully acknowledged.

\newpage


\begin{thebibliography}{99}
\bibitem{x7} 
That is, all laws applying to physical systems where the curvature of
space-time may be neglected, so that General Relativistic effects are unimportant,
and may be neglected.
\bibitem{x14}
The positive sign for $\gamma$ is taken in solving Eqn.(2.12). Evidently
$\gamma \rightarrow 1$ as $\beta \rightarrow 0$.
\bibitem{x21}
It is often stated in the literature that the potentials $\phi$, $\vec{A}$
are introduced only for `reasons of mathematical simplicity' and `have no
physical meaning'. See for example: F.R\"{o}hrlich
 {\it Classical Charged Particles}~(Addison-Wesley 1990), p65-66.
Actually, the underlying space-time
symmetries of Maxwell's equations can only be expressed by using the 
4-vector character of $A^{\mu}$. Also the minimal electromagnetic interaction
 in the covariant formulation
 of relativistic quantum mechanics, which is the
dynamical basis of Quantum Electrodynamics, requires the introduction of
a quantum field for the photon that has a the same 4-vector nature as the
 electromagnetic potential.
\end{thebibliography}
\end{document}